\documentclass[conference]{IEEEtran}
\usepackage{cite}
\usepackage{amsmath,amssymb,amsfonts}
\usepackage{algorithmic}
\usepackage{graphicx}
\usepackage{textcomp}
\usepackage{xcolor}
\usepackage{soul}
\usepackage{subfig}
\usepackage{multirow}
\usepackage{tabulary}
\usepackage[scientific-notation=true]{siunitx}

\usepackage{balance}

\usepackage{color}

\begin{document}

\title{Knowledge Distillation for Efficient Audio-Visual Video Captioning}

\author{\IEEEauthorblockN{
Özkan Çaylı$^1$,
Xubo Liu$^2$,
Volkan Kılıç$^1$\IEEEauthorrefmark{1},
Wenwu Wang$^2$\IEEEauthorrefmark{2}}
\IEEEauthorblockA{$^1$Electrical and Electronics Engineering, İzmir Katip Çelebi University, Türkiye}
\IEEEauthorblockA{$^2$Centre for Vision, Speech and Signal Processing (CVSSP), University of Surrey, UK}
\IEEEauthorblockA{Email: \IEEEauthorrefmark{1}volkan.kilic@ikcu.edu.tr; \IEEEauthorrefmark{2}w.wang@surrey.ac.uk}
}

\maketitle
\begin{abstract}
Automatically describing audio-visual content with texts, namely video captioning, has received significant attention due to its potential applications across diverse fields. 
Deep neural networks are the dominant methods, offering state-of-the-art performance. 
However, these methods are often undeployable in low-power devices like smartphones due to the large size of the model parameters. 
In this paper, we propose to exploit simple pooling front-end and down-sampling algorithms with knowledge distillation for audio and visual attributes using a reduced number of audio-visual frames.
With the help of knowledge distillation from the teacher model, our proposed method greatly reduces the redundant information in audio-visual streams without losing critical contexts for caption generation.
Extensive experimental evaluations on the MSR-VTT dataset demonstrate that our proposed approach significantly reduces the inference time by about 80\% with a small sacrifice (less than 0.02\%) in captioning accuracy.
\end{abstract}


\section{Introduction}
\label{sec:intro}

Video captioning aims to generate grammatically and semantically meaningful sentences for the content of audio-visual media, driven by applications such as video indexing or retrieval and virtual assistants for visually and hearing-impaired people   \cite{uslu2022resnet, fetiler2021video}. 

This task involves several challenges, such as identifying objects and scenes in the video frame, extracting audio attributes, and audio-visual fusion to describe the content with certain grammatical structures and semantics \cite{keskin2021benchmark, mei2022automated, sun2022automated, liu2022visually}.  
These issues could be addressed with the release of large-scale datasets and advances in deep learning, which has led to the development of highly complex networks with improved caption generation.
However, this can also lead to high computational cost due to the increased complexity of the networks and scale of the datasets.
One approach to overcome this issue is to use efficient audio and visual feature extraction networks as they provide faster inference time \cite{howard2017mobilenets}.
These networks can be categorized into four classes: namely, model compression \cite{singh2022passive, singh2022low}, knowledge distillation \cite{futami2020distilling,lu2017knowledge,choi2022temporal}, efficient networks \cite{kong2020panns, liang2020channel}, and simple pooling front-ends (SimPFs) \cite{liu2023simple}. A framework that applies passive filter pruning to reduce the number of convolutional filters is proposed for a compressed convolutional neural network (CNN) \cite{singh2022passive}. 
Similarly, a low-complexity CNN architecture is presented in \cite{singh2022low}, by reducing model parameters and memory usage.
A BERT architecture is proposed as a teacher network that provides soft labels to guide a seq2seq network for audio speech recognition \cite{futami2020distilling}.
In a highway deep neural network, knowledge distillation and teacher-student training are leveraged to achieve improved accuracy with a reduced number of parameters \cite{lu2017knowledge}.
Pretrained audio neural networks (PANNs) \cite{kong2020panns}, which are trained on AudioSet \cite{gemmeke2017audio},  can be transferred to audio-related tasks such as audio classification and captioning \cite{liu2022leveraging, mei2022diverse, mei2021encoder}.
SimPFs are employed to reduce the required number of audio frames by reducing floating point operations on a network for efficient audio classification \cite{liu2023simple}.

For visual feature extraction, knowledge distillation is used in \cite{ostyakov2018label} to generate soft labels for simpler networks to be deployed on a device with low computing resources.
Similarly, knowledge distillation with an attention mechanism is used in \cite{lin2018nextvlad}, which groups high-dimensional features into low-dimensional vectors.
Furthermore, \cite{bhardwaj2019efficient} uses all the visual frames in a video to train the teacher network. The student network then uses uniformly down-sampled frames and mimics the teacher for efficient video classification.

In this study, we propose an efficient audio-visual captioning method based on the teacher-student network, which uses knowledge distillation for audio and visual feature extraction with a reduced number of frames, leading to substantially improved captioning efficiency.
More specifically, the PANNs network \cite{kong2020panns} is used with SimPF \cite{liu2023simple} for audio feature extraction, while Inception-v3 CNN architecture \cite{ioffe2015batch} with down-sampling is utilized for visual feature extraction \cite{ccayli2020mobile}.
The language model uses simple stacked gated recurrent units (GRUs) \cite{ccayli2022auxiliary} with dropouts \cite{wager2013dropout} and residual connections \cite{karpathy2015deep, aydinsequence}.
The student network is first trained and fine-tuned with the cross-entropy loss. To further improve the captioning accuracy, the representation loss is also used along with the cross-entropy loss.
The experiments show that knowledge distillation can speed up audio-visual feature extraction with a negligible drop in captioning accuracy.

This paper is organized as follows: Section \ref{sec:method} presents the proposed audio-visual video captioning approach. 
Section \ref{sec:eval} describes the dataset and performance metrics for experimental evaluations and discusses the experimental results, followed by the conclusions.

\section{Proposed Approach}
\label{sec:method}

\begin{figure*}[t]
    \centering
    \includegraphics[width=\linewidth]{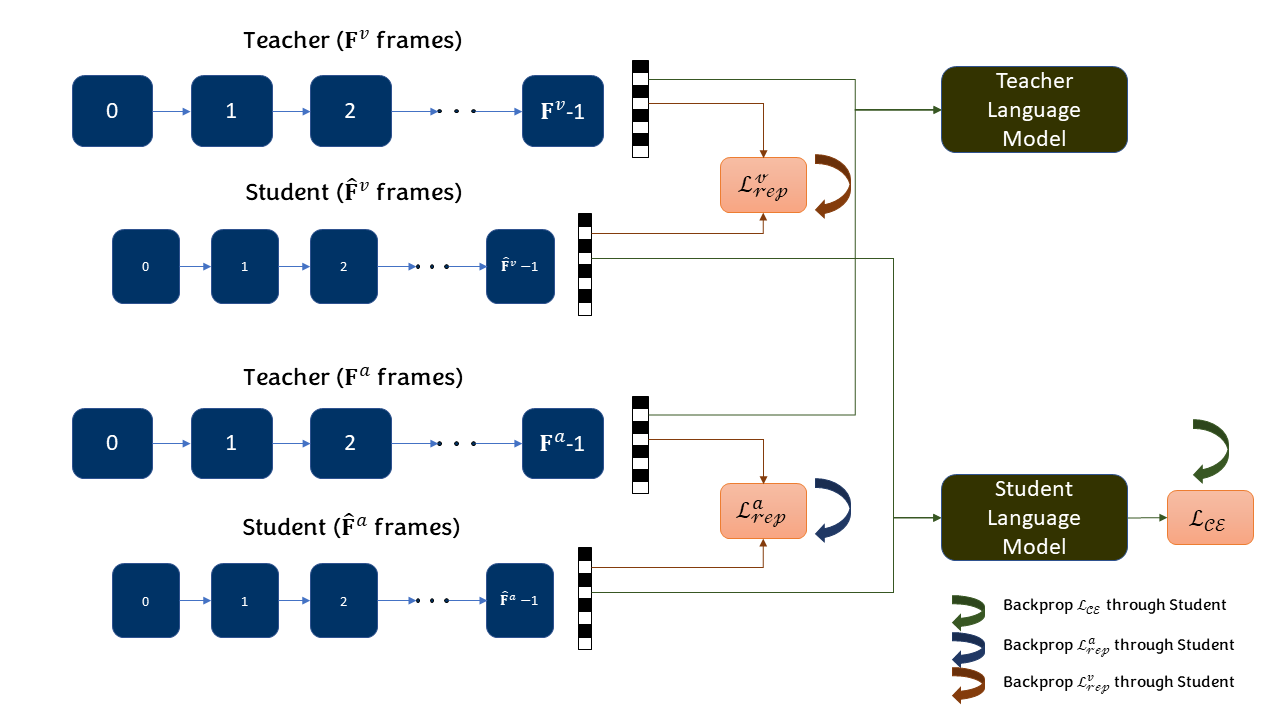}
    \caption{The Proposed Approach}
    \label{fig:modelFigure}
\end{figure*}

This section presents the proposed video captioning approach based on the teacher-student model, as illustrated in Figure \ref{fig:modelFigure}. 

In video captioning, a sequence of words needs to be predicted from a vocabulary using audio and visual attributes. 
The teacher network utilizes $N^a$ audio frames ${\bf F}^a=(F^a_0,F^a_{1},...F^a_{N_a-1})$ and $N^v$ video frames $ {\bf F}^v=(F^v_0,F^v_1,...F^v_{N_v-1})$ of the video $\textbf{V}$  to predict a caption which can be stated using a neural network $f$:

\begin{equation}
\label{eq:avc}
P(\hat{\textbf{Y}}|\textbf{V}) = f({\bf F}^a ,{\bf F}^v),    
\end{equation}  
where $\hat{\textbf{Y}}$ denotes a series of words as $(\hat{y}_{0}, \hat{y}_{1}, ... \hat{y}_{N^c})$ and $N^c$ refers to the number of words in the caption.

We employ Inception-v3 CNN architecture pre-trained on the ImageNet dataset to extract features from visual frames.
The architecture resizes the images to 3$\times$299$\times$299, then the average pooling layer outputs a latent vector consisting of 2048 units.
Similarly, audio features are extracted with PANNs CNN architecture containing 10 stacked CNN layers pre-trained on AudioSet.
A recurrent neural network (RNN)-based network that utilizes audio and visual features from the Inception-v3 and PANNs is used as a language model to generate captions.
We employ a mean operator and acquire latent vectors from time-series input, which describe audio and visual features. 
These latent vectors are concatenated and fed to the RNN-based network consisting of embedding, GRUs, and linear layers.
Moreover, residual connections and dropouts are applied between layers to maintain gradient flow from the lower to upper layers.
The teacher network is trained with the cross-entropy loss denoted as $\mathcal{L}_{CE}$. 
The student network is similar to the teacher, where SimPF and down-sampling algorithms are employed to reduce the number of audio and visual frames by a compression rate in a video. Specifically, we use the spectral pooling method of SimPF, which computes the discrete Fourier transform (DFT) of the audio frames ${\bf F}^a$ and then crops the center with a bounding box with the shape of $(S, kN^a)$ where $S$ refers to the dimension of the spectral feature to get $\tilde{{\bf F}}^a_{crop}$.
\begin{table*}[]
\caption{Performance metric evaluation results on the MSR-VTT test set}
\label{tab:metric-results_msrvtt_test}
\setlength{\tabcolsep}{6pt}
\renewcommand{\arraystretch}{1.5}
\centering
\begin{tabular}{|l|cccccccc|c|c|}

\hline

         & BLEU-1 & BLEU-2 & BLEU-3 & BLEU-4 & CIDEr & METEOR & ROUGE-L & SPICE & SCORE & Diff (\%) \\ \hline

  $\mathcal{L}_{rep}$ Student (k = 0.2) & 0.722 & 0.555 & 0.422 & 0.311 & 0.267  & 0.236 & 0.554 & 0.045 & 0.321 & 0.127 \\ \hline
  $\mathcal{L}_{rep}$ Student (k = 0.4) & 0.715 & 0.546 & 0.411 & 0.294 & 0.223  & 0.234 & 0.539 & 0.043 & 0.306 & 0.168 \\ \hline
  $\mathcal{L}_{rep}$ Student (k = 0.6) & 0.709 & 0.542 & 0.412 & 0.300 & 0.232  & 0.231 & 0.543 & 0.041 & 0.308 & 0.163 \\ \hline
  $\mathcal{L}_{rep}$ Student (k = 0.8) & 0.719 & 0.550 & 0.413 & 0.300 & 0.256  & 0.235 & 0.545 & 0.046 & 0.315 & 0.144 \\ \hline
  $\mathcal{L}_{rep}+\mathcal{L}_{CE}$ Student (k = 0.2) & 0.766 & 0.613 & 0.476 & 0.357 & 0.375  & 0.256 & 0.585 & 0.054 & 0.365 & 0.008 \\ \hline
  $\mathcal{L}_{rep}+\mathcal{L}_{CE}$ Student (k = 0.4) & \textbf{0.774} & \textbf{0.618} & 0.473 & 0.348 & 0.359  & 0.256 & 0.582 & \textbf{0.055} & 0.361 & 0.019 \\ \hline
  $\mathcal{L}_{rep}+\mathcal{L}_{CE}$ Student (k = 0.6) & 0.769 & 0.616 & 0.478 & 0.357 & 0.375  & \textbf{0.258} & \textbf{0.586} & \textbf{0.055} & 0.366 & 0.005 \\ \hline
  $\mathcal{L}_{rep}+\mathcal{L}_{CE}$ Student (k = 0.8) & 0.765 & 0.614 & \textbf{0.479} & \textbf{0.358} & 0.366  & 0.255 & 0.583 & 0.054 & 0.362 & 0.016 \\ \hline
  Teacher & 0.760 & 0.612 & 0.473 & 0.352 &  \textbf{0.397} & 0.254  & 0.583 & 0.054 & \textbf{0.368} & 0.000 \\ \hline
\end{tabular}
\end{table*}
Then the output of the inverse discrete Fourier transform (IDFT) $\hat{{\bf F}}^a$ is taken as the compressed audio, as shown below, 

\begin{equation}
    \label{eq:ssimpf}
    \begin{split}
    \tilde{{\bf F}}^a &= DFT({\bf F}^a) \\  
    \tilde{{\bf F}}^a_{crop} &= {\bf F}^a (S, kN^a) \\
    \hat{{\bf F}}^a &= IDFT(\tilde{{\bf F}}^a_{crop}).
    \end{split}
\end{equation}

Down-sampling is performed on ${\bf F}^v$ to obtain compressed visual frames $\hat{\bf F}^v$,

\begin{equation}
    \label{eq:down-sample}
    \hat{\bf F}^v = {\bf F}^v(m/k), \qquad \qquad
    m = 0, 1, 2, ..., N_v-1
\end{equation}
where $k$ denotes the compression rate, ranging from 0 to 1.

We extract audio and visual features from compressed frames using PANNs and Inception-v3. Then, latent vectors are acquired with a mean operator. We employ knowledge distillation from the teacher network to increase the accuracy of caption generation. A neural network with two hidden layers is utilized to increase the resemblance of latent vectors to the teacher. The network is trained to minimize the L1 loss between student and teacher latent vectors. We denote this loss as $\mathcal{L}_{rep}$ where rep refers to representation. We train the teacher network, and then the teacher guides the optimization of the parameters of the student network. In this study, we train the student-teacher network with the following losses:

$\mathcal{L}_{rep}$: The student network is only trained by the $\mathcal{L}_{rep}$ loss and is learned to mimic the audio-visual features of the teacher network. Then, the language model is trained with the updated neural network.

$\mathcal{L}_{rep} + \mathcal{L}_{CE}$: 
We employ both $\mathcal{L}_{rep}$ and $\mathcal{L}_{CE}$ losses to minimize the representation loss and maximize the captioning accuracy.

\section{Experimental Evaluations}
\label{sec:eval}

\subsection{Setup and Performance Metrics}

The proposed approach is evaluated on the MSR-VTT dataset \cite{xu2016msr}, which initially consists of 10,000 videos, each with 20 ground-truth captions. 
However, by the time the experiments are executed, only 5,074 and 2,123 videos are available from the training and testing sets, respectively.
Several performance metrics are employed to measure the accuracy of the video captioning approach, 
including
metrics for evaluation of translation with explicit ordering (METEOR) \cite{lavie2007meteor}, bilingual evaluation understudy (BLEU) \cite{papineni2002bleu}, consensus-based image description evaluation (CIDEr) \cite{vedantam2015cider}, and recall-oriented understudy for gisting evaluation-longest common subsequence (ROUGE-L) \cite{lin2004rouge}, and semantic propositional image caption evaluation (SPICE) \cite{anderson2016spice}.

The ranking of the results is based on a final SCORE which is calculated as an average of all performance metrics. In calculating the final SCORE, we used the mean of the BLEU scores.
For the experiments, the visual frames of the videos are resized into the shape of 3$\times$299$\times$299. 
We utilized tokenization and punctuation removal on the ground-truth captions of the training set. The latent vector size of the layers in the language models is set to 2,576, and the dimension of the linear layer output is equal to the vocabulary length. We evaluated the proposed approach with $0.2, 0.4, 0.6,$ and $ 0.8$ compression ratios.

\subsection{Results \& Discussion}

The accuracy and time consumption of the teacher and student networks are measured with the test set of the MSR-VTT dataset under the $\mathcal{L}_{rep}$, and $\mathcal{L}_{rep} + \mathcal{L}_{CE}$ losses.
In the evaluations, we compressed the frames on the student networks to enable faster inference time. 
The results for the students and teacher networks are given in Table \ref{tab:metric-results_msrvtt_test}, while time consumptions are shown in Table \ref{tab:time-consumption-results_msrvtt_test}.

 Using only the $\mathcal{L}_{rep}$ loss resulted in poor captioning performance in all performance metrics regarding the teacher network, as seen in Table \ref{tab:metric-results_msrvtt_test}. Notably, among the student networks trained with the $\mathcal{L}_{rep}$ loss, the compression rate of $0.2$ has achieved the highest final SCORE. However, the combination of the $\mathcal{L}_{rep}$ and $\mathcal{L}_{CE}$ losses in the student networks offered an accuracy approaching the level of the teacher model across all performance metrics. 

\begin{table}[]
\caption{Time consumption evaluation results on random 100 videos from the MSR-VTT test set}
\label{tab:time-consumption-results_msrvtt_test}
\setlength{\tabcolsep}{10pt}
\renewcommand{\arraystretch}{1.5}
\centering
\begin{tabular}{|c|c|c|}
\hline
Network & average time consumption (s) & Diff (\%) \\ \hline
Student (k = 0.2) & 2.77 &  79.1 \\ \hline
Student (k = 0.4) & 5.65 & 57.4 \\ \hline
Student (k = 0.6) & 8.31 & 37.4 \\  \hline
Student (k = 0.8) & 11.03 & 16.9 \\ \hline
Teacher & 13.28 & 0.0 \\ \hline

\end{tabular}
\end{table}

The captioning accuracy of the student network is increased from $0.321$ to $0.365$  with $\mathcal{L}_{rep}$+$\mathcal{L}_{CE}$ under the compression rate of $0.2$. 
The difference between the accuracy of the teacher and student network dropped from $0.127\%$ to $0.008\%$.
However, the student network with a $0.4$ compression rate leveraged the final SCORE from $0.306$ to $0.361$, which is still lower than that of the compression rate at $0.2$. 
We achieved the highest final SCORE at $0.366$ using the student network with a compression rate of $0.6$. This is followed by the compressed student network with a compression rate of $0.8$, with a final SCORE of $0.362$.
Furthermore, the student networks with compressed audio and visual frames scored higher across some metrics than the teacher. This indicates that student networks can generate accurate captions similar to the teacher. In Table \ref{tab:time-consumption-results_msrvtt_test}, we present the time consumption of feature extraction for both audio and visual frames from randomly selected $100$ videos from the test set of the MSR-VTT dataset. The compression rate $0.8$ reduces feature extraction time up to $16.9\%$, while $0.6$ compression rate decreases the audio-visual feature extraction time by about $37.4\%$. Similarly, $0.4$ and $0.2$ have reduced the inference time by $57.4\%$ and $79.1\%$, respectively. Table \ref{tab:time-consumption-results_msrvtt_test} shows that the student networks reduce inference time significantly compared to the teacher network.

\section{Conclusion}  
\label{sec:conclusion} 

In this study, we have presented a simple pooling front-end and down-sampling method to reduce the number of audio and visual frames in a video for video captioning. Furthermore, we have proposed a teacher-student based-network to leverage the accuracy of caption generation with knowledge distillation. We used $\mathcal{L}_{rep}$ representation and $\mathcal{L}_{CE}$ cross-entropy loss for network training. The proposed approach is evaluated on the MSR-VTT dataset.
Experimental results show that the proposed approach significantly reduces the inference time with a negligible drop in captioning accuracy.

\section*{Acknowledgment} 
This research was supported by the Scientific and Technological Research Council of Turkey (TUBITAK)-British Council (The Newton-Katip Celebi Fund Institutional Links, Turkey-UK projects: 120N995, \& 623805725) and by the scientific research projects coordination unit of Izmir Katip Celebi University (project no: 2021-ÖDL-MÜMF-0006, \& 2022-TYL-FEBE-0012). For the purpose of open access, the author has applied a Creative Commons Attribution (CC BY) licence to any Author Accepted Manuscript version arising.

\bibliographystyle{IEEEbib}
\bibliography{my_reference}

\begin{thebibliography}{10}

\bibitem{uslu2022resnet}
Bet{\"u}l Uslu, {\"O}zkan {\c{C}}ayl{\i}, Volkan K{\i}l{\i}{\c{c}}, and
  Aytu{\u{g}} Onan,
\newblock ``Resnet based deep gated recurrent unit for image captioning on
  smartphone,''
\newblock {\em European Journal of Science and Technology}, , no. 35, pp.
  610--615, 2022.

\bibitem{fetiler2021video}
Beng{\"u} Fetiler, {\"O}zkan Çaylı, {\"O}zge~Taylan Moral, Volkan Kılıç,
  and Aytu{\u{g}} Onan,
\newblock ``Video captioning based on multi-layer gated recurrent unit for
  smartphones,''
\newblock {\em European Journal of Science and Technology}, , no. 32, pp.
  221--226, 2021.

\bibitem{keskin2021benchmark}
Rumeysa Keskin, {\"O}zkan Çaylı, {\"O}zge~Taylan Moral, Volkan Kılıç, and
  Aytu{\u{g}} Onan,
\newblock ``A benchmark for feature-injection architectures in image
  captioning,''
\newblock {\em European Journal of Science and Technology}, , no. 31, pp.
  461--468, 2021.

\bibitem{mei2022automated}
Xinhao Mei, Xubo Liu, Mark~D Plumbley, and Wenwu Wang,
\newblock ``Automated audio captioning: An overview of recent progress and new
  challenges,''
\newblock {\em EURASIP journal on audio, speech, and music processing}, vol.
  2022, no. 1, pp. 1--18, 2022.

\bibitem{sun2022automated}
Jianyuan Sun, Xubo Liu, Xinhao Mei, Mark~D Plumbley, Volkan Kilic, and Wenwu
  Wang,
\newblock ``Automated audio captioning via fusion of low-and high-dimensional
  features,''
\newblock {\em arXiv preprint arXiv:2210.05037}, 2022.

\bibitem{liu2022visually}
Xubo Liu, Qiushi Huang, Xinhao Mei, Haohe Liu, Qiuqiang Kong, Jianyuan Sun,
  Shengchen Li, Tom Ko, Yu~Zhang, Lilian~H Tang, et~al.,
\newblock ``Visually-aware audio captioning with adaptive audio-visual
  attention,''
\newblock {\em arXiv preprint arXiv:2210.16428}, 2022.

\bibitem{howard2017mobilenets}
Andrew~G Howard, Menglong Zhu, Bo~Chen, Dmitry Kalenichenko, Weijun Wang,
  Tobias Weyand, Marco Andreetto, and Hartwig Adam,
\newblock ``Mobilenets: Efficient convolutional neural networks for mobile
  vision applications,''
\newblock {\em arXiv preprint arXiv:1704.04861}, 2017.

\bibitem{singh2022passive}
Arshdeep Singh and Mark~D Plumbley,
\newblock ``A passive similarity based cnn filter pruning for efficient
  acoustic scene classification,''
\newblock {\em arXiv preprint arXiv:2203.15751}, 2022.

\bibitem{singh2022low}
Arshdeep Singh and Mark~D Plumbley,
\newblock ``Low-complexity cnns for acoustic scene classification,''
\newblock {\em arXiv preprint arXiv:2207.11529}, 2022.

\bibitem{futami2020distilling}
Hayato Futami, Hirofumi Inaguma, Sei Ueno, Masato Mimura, Shinsuke Sakai, and
  Tatsuya Kawahara,
\newblock ``Distilling the knowledge of bert for sequence-to-sequence asr,''
\newblock {\em arXiv preprint arXiv:2008.03822}, 2020.

\bibitem{lu2017knowledge}
Liang Lu, Michelle Guo, and Steve Renals,
\newblock ``Knowledge distillation for small-footprint highway networks,''
\newblock in {\em International Conference on Acoustics, Speech and Signal
  Processing (ICASSP)}. IEEE, 2017, pp. 4820--4824.

\bibitem{choi2022temporal}
Kwanghee Choi, Martin Kersner, Jacob Morton, and Buru Chang,
\newblock ``Temporal knowledge distillation for on-device audio
  classification,''
\newblock in {\em International Conference on Acoustics, Speech and Signal
  Processing (ICASSP)}. IEEE, 2022, pp. 486--490.

\bibitem{kong2020panns}
Qiuqiang Kong, Yin Cao, Turab Iqbal, Yuxuan Wang, Wenwu Wang, and Mark~D
  Plumbley,
\newblock ``Panns: Large-scale pretrained audio neural networks for audio
  pattern recognition,''
\newblock {\em Transactions on Audio, Speech, and Language Processing}, vol.
  28, pp. 2880--2894, 2020.

\bibitem{liang2020channel}
Jinhua Liang, Tao Zhang, and Guoqing Feng,
\newblock ``Channel compression: Rethinking information redundancy among
  channels in cnn architecture,''
\newblock {\em IEEE Access}, vol. 8, pp. 147265--147274, 2020.

\bibitem{liu2023simple}
Xubo Liu, Haohe Liu, Qiuqiang Kong, Xinhao Mei, Mark~D Plumbley, and Wenwu
  Wang,
\newblock ``Simple pooling front-ends for efficient audio classification,''
\newblock in {\em IEEE International Conference on Acoustics, Speech and Signal
  Processing (ICASSP)}. IEEE, 2023, pp. 1--5.

\bibitem{gemmeke2017audio}
Jort~F Gemmeke, Daniel~PW Ellis, Dylan Freedman, Aren Jansen, Wade Lawrence,
  R~Channing Moore, Manoj Plakal, and Marvin Ritter,
\newblock ``Audio set: An ontology and human-labeled dataset for audio
  events,''
\newblock in {\em 2017 IEEE international conference on acoustics, speech and
  signal processing (ICASSP)}. IEEE, 2017, pp. 776--780.

\bibitem{liu2022leveraging}
Xubo Liu, Xinhao Mei, Qiushi Huang, Jianyuan Sun, Jinzheng Zhao, Haohe Liu,
  Mark~D Plumbley, Volkan Kilic, and Wenwu Wang,
\newblock ``Leveraging pre-trained bert for audio captioning,''
\newblock in {\em 2022 30th European Signal Processing Conference (EUSIPCO)}.
  IEEE, 2022, pp. 1145--1149.

\bibitem{mei2022diverse}
Xinhao Mei, Xubo Liu, Jianyuan Sun, Mark~D Plumbley, and Wenwu Wang,
\newblock ``Diverse audio captioning via adversarial training,''
\newblock in {\em IEEE International Conference on Acoustics, Speech and Signal
  Processing (ICASSP)}. IEEE, 2022, pp. 8882--8886.

\bibitem{mei2021encoder}
Xinhao Mei, Qiushi Huang, Xubo Liu, Gengyun Chen, Jingqian Wu, Yusong Wu,
  Jinzheng Zhao, Shengchen Li, Tom Ko, H~Lilian Tang, et~al.,
\newblock ``An encoder-decoder based audio captioning system with transfer and
  reinforcement learning,''
\newblock {\em arXiv preprint arXiv:2108.02752}, 2021.

\bibitem{ostyakov2018label}
Pavel Ostyakov, Elizaveta Logacheva, Roman Suvorov, Vladimir Aliev, Gleb
  Sterkin, Oleg Khomenko, and Sergey~I. Nikolenko,
\newblock ``Label denoising with large ensembles of heterogeneous neural
  networks,''
\newblock in {\em Proceedings of the European Conference on Computer Vision
  (ECCV) Workshops}, 2018, pp. 250--261.

\bibitem{lin2018nextvlad}
Rongcheng Lin, Jing Xiao, and Jianping Fan,
\newblock ``Nextvlad: An efficient neural network to aggregate frame-level
  features for large-scale video classification,''
\newblock in {\em Proceedings of the European Conference on Computer Vision
  (ECCV) Workshops}, 2018, pp. 1092--1101.

\bibitem{bhardwaj2019efficient}
Shweta Bhardwaj, Mukundhan Srinivasan, and Mitesh~M Khapra,
\newblock ``Efficient video classification using fewer frames,''
\newblock in {\em Proceedings of the Conference on Computer Vision and Pattern
  Recognition (CVF)}. 2019, pp. 354--363, IEEE.

\bibitem{ioffe2015batch}
Sergey Ioffe and Christian Szegedy,
\newblock ``Batch normalization: Accelerating deep network training by reducing
  internal covariate shift,''
\newblock in {\em International Conference on Machine Learning}. PMLR, 2015,
  pp. 448--456.

\bibitem{ccayli2020mobile}
{\"O}zkan {\c{C}}ayl{\i}, Burak Makav, Volkan K{\i}l{\i}{\c{c}}, and
  Aytu{\u{g}} Onan,
\newblock ``Mobile application based automatic caption generation for visually
  impaired,''
\newblock in {\em International Conference on Intelligent and Fuzzy Systems
  (INFUS)}. IEEE, 2020, pp. 1532--1539.

\bibitem{ccayli2022auxiliary}
{\"O}zkan {\c{C}}ayl{\i}, Volkan K{\i}l{\i}{\c{c}}, Aytu{\u{g}} Onan, and Wenwu
  Wang,
\newblock ``Auxiliary classifier based residual rnn for image captioning,''
\newblock in {\em 2022 30th European Signal Processing Conference (EUSIPCO)}.
  IEEE, 2022, pp. 1126--1130.

\bibitem{wager2013dropout}
Stefan Wager, Sida Wang, and Percy~S Liang,
\newblock ``Dropout training as adaptive regularization,''
\newblock {\em Advances in Neural Information Processing Systems}, vol. 26,
  2013.

\bibitem{karpathy2015deep}
Andrej Karpathy and Li~Fei-Fei,
\newblock ``Deep visual-semantic alignments for generating image
  descriptions,''
\newblock in {\em Proceedings of the Conference on Computer Vision and Pattern
  Recognition (CVF)}. 2015, pp. 3128--3137, IEEE.

\bibitem{aydinsequence}
Selman Aydın, {\"O}zkan {\c{C}}aylı, Volkan Kılı{\c{c}}, and Aytu{\u{g}}
  Onan,
\newblock ``Sequence-to-sequence video captioning with residual connected gated
  recurrent units,''
\newblock {\em European Journal of Science and Technology}, , no. 35, pp.
  380--386, 2022.

\bibitem{xu2016msr}
Jun Xu, Tao Mei, Ting Yao, and Yong Rui,
\newblock ``Msr-vtt: A large video description dataset for bridging video and
  language,''
\newblock in {\em Proceedings of the Conference on Computer Vision and Pattern
  Recognition (CVF)}. 2016, pp. 5288--5296, IEEE.

\bibitem{lavie2007meteor}
Alon Lavie and Abhaya Agarwal,
\newblock ``Meteor: An automatic metric for mt evaluation with high levels of
  correlation with human judgments,''
\newblock in {\em Proceedings of the Second Workshop on Statistical Machine
  Translation}, 2007, pp. 228--231.

\bibitem{papineni2002bleu}
Kishore Papineni, Salim Roukos, Todd Ward, and Wei-Jing Zhu,
\newblock ``Bleu: a method for automatic evaluation of machine translation,''
\newblock in {\em Proceedings of the 40th Annual Meeting on Association for
  Computational Linguistics (ACL)}, 2002, pp. 311--318.

\bibitem{vedantam2015cider}
Ramakrishna Vedantam, C~Lawrence~Zitnick, and Devi Parikh,
\newblock ``Cider: Consensus-based image description evaluation,''
\newblock in {\em Proceedings of the Conference on Computer Vision and Pattern
  Recognition (CVF)}. 2015, pp. 4566--4575, IEEE.

\bibitem{lin2004rouge}
Chin-Yew Lin,
\newblock ``Rouge: A package for automatic evaluation of summaries,''
\newblock in {\em Proceedings of the Association for Computational Linguistics
  (ACL) Workshop}, 2004, pp. 1--8.

\bibitem{anderson2016spice}
Peter Anderson, Basura Fernando, Mark Johnson, and Stephen Gould,
\newblock ``Spice: Semantic propositional image caption evaluation,''
\newblock in {\em European Conference on Computer Vision (ECCV)}. Springer,
  2016, pp. 382--398.

\end{thebibliography}

\end{document}